\documentclass{article}
\usepackage[preprint]{spconf}
                \copyrightnotice{\copyright\ IEEE 2019}
                \toappear{To appear in {\it Proc.\ ICASSP 2019, May 12-17, 2019, Brighton, UK}}
\usepackage{spconf}
\usepackage{amsmath,graphicx,url}
\usepackage{multirow}

\title{When CTC Training Meets Acoustic Landmarks}
\name{Di He$^{\ast,1,2}$\thanks{$^\ast$Authors contributed equally.}, Xuesong Yang$^{\ast,1,3}$, Boon Pang Lim$^2$, Yi Liang$^1$, Mark Hasegawa-Johnson$^1$, Deming Chen$^1$}
\address{$^1$ECE, Coordinated Science Lab, and Beckman Institute, University of Illinois, Urbana, IL, USA\\
$^2$Novumind Inc, Santa Clara, CA, USA \qquad $^3$Amazon Alexa Speech, Seattle, WA, USA}
\begin{document}
%
\maketitle
\begin{abstract}
Connectionist temporal classification (CTC) provides an end-to-end acoustic model (AM) training strategy. CTC learns accurate AMs without time-aligned phonetic transcription, but sometimes fails to converge, especially in resource-constrained scenarios. In this paper, the convergence properties of CTC are improved by incorporating acoustic landmarks. We tailored a new set of acoustic landmarks to help CTC training converge more rapidly and smoothly while also reducing recognition error rates. We leveraged new target label sequences mixed with both phone and manner changes to guide CTC training. Experiments on TIMIT demonstrated that CTC based acoustic models converge significantly faster and smoother when they are augmented by acoustic landmarks. The models pretrained with mixed target labels can be further finetuned, resulting in phone error rates $8.72\%$ below baseline on TIMIT. Consistent performance gain is also observed on WSJ (a larger corpus) and reduced TIMIT (smaller). With WSJ, we are the first to succeed in verifying the effectiveness of acoustic landmark theory on a mid-sized ASR task. 
\end{abstract}
\begin{keywords}
Acoustic Modeling, CTC, Acoustic Landmarks, End-to-End
\end{keywords}
\section{Introduction}
\label{sec:intro}
Automatic speech recognition (ASR) is a sequence labeling problem that translates a speech waveform into a sequence of words. Recent success of hidden Markov model (HMM) combined with deep neural networks (DNNs) or recurrent neural networks has achieved a word error rate (WER) on par with human transcribers~\cite{xiong2016achieving,saon2017english}. These hybrid acoustic models (AMs) are typically optimized by cross-entropy (CE) training which relies on accurate frame-wise context-dependent state alignments pre-generated from a seed AM. The connectionist temporal classification (CTC) loss function~\cite{graves2006connectionist}, in contrast, provides an alternative method of AM training in an end-to-end fashion---it directly addresses the sequence labeling problem without prior frame-wise alignments. CTC is capable of learning to construct frame-wise paths implicitly bridging between the input speech waveform and its context-independent target, and it has been demonstrated to outperform hybrid HMM systems when the amount of training data is large~\cite{sak2015fast,miao2015eesen,amodei2016deep}. However, its performance degrades and is even worse than traditional CE training when applied to small-scale data~\cite{miao2016empirical}. 

Training CTC models can be time-consuming and sometimes models are apt to converge to even a sub-optimal alignment, especially on resource-constrained data. In order to alleviate such common problems of CTC training, additional tricks are needed, for example, ordering training utterances by their lengths~\cite{amodei2016deep} or bootstrapping CTC models with models CE-trained on fixed alignments~\cite{sak2015learning}. The success of bootstrapping with prior alignments indicates that external phonetic knowledge may help to regularize CTC training towards stable and fast convergence. Furthermore, another investigation~\cite{niu2017study} reveals that the spiky predictions of CTC models tend to overlap with the vicinity of acoustic landmarks where abrupt manner changes of articulation occur~\cite{stevens2002toward}. 
The possible coincidence of CTC peaks overlapping acoustic landmarks suggests a number of possible approaches for reducing the data requirements of CTC, including cross-language transfer (using the relative language-independence of acoustic landmarks~\cite{he2018improved}) and informative priors. 


Many efforts have been attempted to augment acoustic modeling with acoustic landmarks~\cite{he2018improved,he2017selecting,he2018acoustic} which are detected by accurate time-aligned phonetic transcriptions. To the best of our knowledge, only TIMIT~\cite{garofalo1993darpa} (5.4 hours) provides such fine-grained transcriptions. The value of testing these approaches are limited since the only available corpus is very small. It is worth further exploring the power of landmark theory when scaled up to large corpus speech recognition.  

In this paper, we propose to augment phone sequences with acoustic landmarks for CTC acoustic modeling and 
leverage a two-phase training procedure with pretraining and finetuning to address CTC convergence problems. Experiments on TIMIT demonstrate that our approaches not only help CTC models converge more rapidly and smoothly, but also achieve a lower phone error rate, up to $8.72\%$ phone error rate reduction over CTC baseline with phone labels only. We also investigate the sensitivity of our approaches to the size of training data on subsets of TIMIT (smaller corpora) and WSJ~\cite{paul1992design} (a larger corpus). Our findings demonstrate that label augmentation generalizes to larger and smaller training datasets, and we believe this is the first work that applies acoustic landmark theory to a mid-sized ASR corpus.

\section{Background}
\label{sec:back}
\subsection{Connectionist Temporal Classification (CTC)}

Recent end-to-end systems have attracted much attention, for example, because they avoid time-consuming iterations between alignment and model building~\cite{graves2006connectionist,graves2014towards}. The CTC loss computes the total likelihood of the target label sequence over all possible alignments given an input feature sequence, so that the computation is more expensive than frame-wise cross-entropy training. A blank symbol is introduced to compensate for the difference in length between an input feature sequence and its target label sequence. Forward-backward algorithms are used to efficiently sum the likelihood over all possible alignments. The CTC loss is defined as,
\begin{equation*}\label{eq:1}
    {\mathcal L}_{ctc}=-\log p\left( \mathbf{y} | \mathbf{x} \right)=-\log \sum_{\boldsymbol{\pi} \in \mathcal{B}^{-1}(\mathbf{y})}p(\boldsymbol{\pi} | \mathbf{x})
\end{equation*}
where $\mathbf{x}$ is an input feature sequence, $\mathbf{y}$ is the target label sequence of $\mathbf{x}$, $\boldsymbol{\pi}$ is one of blank-augmented alignments of $\mathbf{y}$, and $\mathcal{B}^{-1}(\mathbf{y})$ calculates the set of all such alignments. 
During decoding, the n-best list of predicted label sequences can be achieved by either a greedy search or a beam search based on weighted finite state transducers (WFSTs). In the following experiments, our acoustic models are trained by the phoneme CTC loss, and we report phone error rates on TIMIT (a smaller corpus) through an one-best greedy search and word error rates on WSJ (a larger corpus) through an one-best WFSTs beam search, respectively. 

\subsection{Acoustic Landmarks}
Acoustic landmark theory originates from experimental studies of human speech production and speech perception. It claims there exist instantaneous acoustic events that are perceptually salient and sufficient to distinguish phonemes~\cite{stevens2002toward}. 
Automatic landmark detectors can be knowledge-based~\cite{liu1996landmark} or learned~\cite{hasegawa2005landmark}. 
Landmark-based ASR has been shown to slightly reduce the WER of a large-vocabulary speech recognizer, but only in a rescoring paradigm
using a very small test set~\cite{hasegawa2005landmark}.
Landmarks can reduce computational load for DNN/HMM hybrid models~\cite{he2017selecting,he2018acoustic} and can improve recognition accuracy~\cite{he2018improved}. 
Previous works~\cite{he2018improved,he2017selecting,he2018acoustic,kong2016landmark} annotated landmark positions mostly following experimental findings presented in~\cite{stevens1992implementation,hasegawa2000time}.
Four different landmarks are defined to capture positions of vowel peak, glide valley in glide-like consonants, oral closure and oral release.
\section{Methods}
\label{sec:meth}

\subsection{Distinctive Features and Landmark Definition}\label{sec:label}
Distinctive features (DFs) concisely describe sounds of a language at a sub-segmental level, and they have direct relations to acoustics and articulation. These features take on binary encodings of perceptual, phonological, and articulatory speech sounds~\cite{stevens1981evidence}. A collection of these binary features can distinguish each segment from all others in a language. Autosegmental phonology~\cite{mccarthy1988feature} also suggests that DFs have an internal organization with a hierarchical relationship with each other. We follow these linguistic rules to select two primary features---\emph{sonorant} and \emph{continuant}---that distinguish among the manner classes of articulation, resulting in a four-way categorization shown in Table~\ref{tab:label_mark}.  We define landmarks to be changes in the value of one of these two distinctive features using the TIMIT phone inventory.
The standard phoneme set used by WSJ ignores detailed annotations of oral closures, for example /bcl/, so that we merge together [-,+\emph{continuant}] features under [-\emph{sonorant}] column in Table~\ref{tab:label_mark}, resulting in a three-way categorization for WSJ experiments instead.
\begin{table}[htbp]
    \centering
    \caption{Broad classes of sounds on TIMIT}
    \label{tab:label_mark}
    \vspace{2mm}
    \small
    \begin{tabular}{|l|l|l|}
    \hline
    Manner & -sonorant & +sonorant \\ 
    \hline
    -continuant & bcl dcl gcl kcl & em en eng m n ng \\
     & pcl q tcl & \\ 
    \hline
    +continuant & b d g k p t ch jh & aa ae ah ao aw ax ax-h \\ 
    & dh f hh hv s sh  & axr ay dx eh el ey ih ix \\
    & th v z zh & iy l nv ow oy r uh uw\\
    & &  ux w y er\\
\hline
\end{tabular}
\end{table}

\subsection{Augmenting Phone Sequences With Landmarks}
\begin{figure}[htbp]
    \centering
    \includegraphics[width=0.9\linewidth]{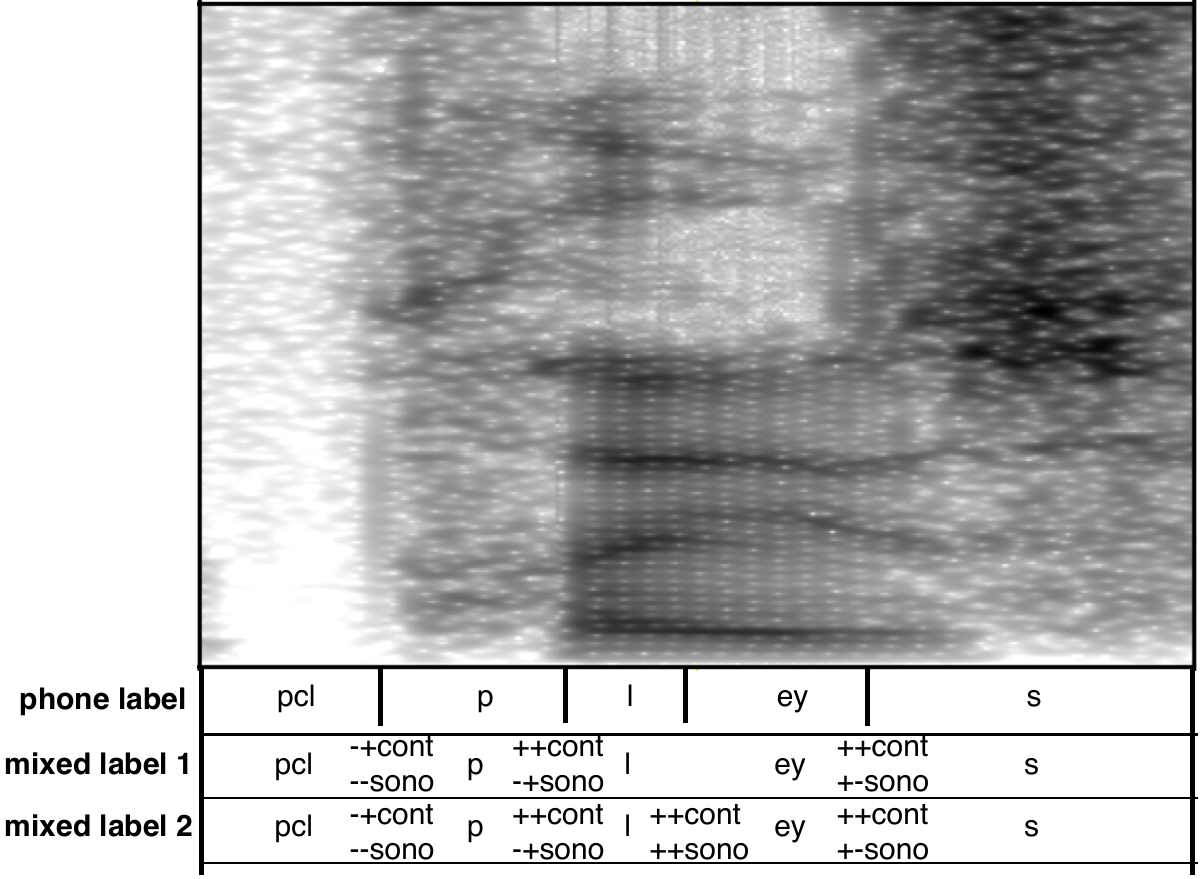}
    \vspace{-4mm}
    \caption{Examples of target label sequences for the word ``PLACE''. The audio clip is selected from $\mathtt{SI792}$ on TIMIT.}
    \label{fig:example_label}
\end{figure}

We defined two methods of augmenting phone label sequences with acoustic landmarks. \emph{Mixed Label 1} only inserts landmarks between two broad classes of sounds where manner changes occur; \emph{Mixed Label 2} inserts landmarks between phones even if manner changes don't exist. Figure~\ref{fig:example_label} demonstrates an example of our two augmentation methods.

CTC only requires a single target label sequence, so that augmenting phone sequences with landmarks can relax the need for time-aligned phone transcriptions. With a blank label present between two phones in the training target sequence, the vanilla CTC training can be considered as already experimenting with the scenario where a dedicated phone boundary label is added to the label set. CTC is thus an ideal baseline for our experiments.

\subsection{Acoustic Modeling using CTC}
\begin{figure}[htbp]
    \centering
    \includegraphics[width=0.85\linewidth]{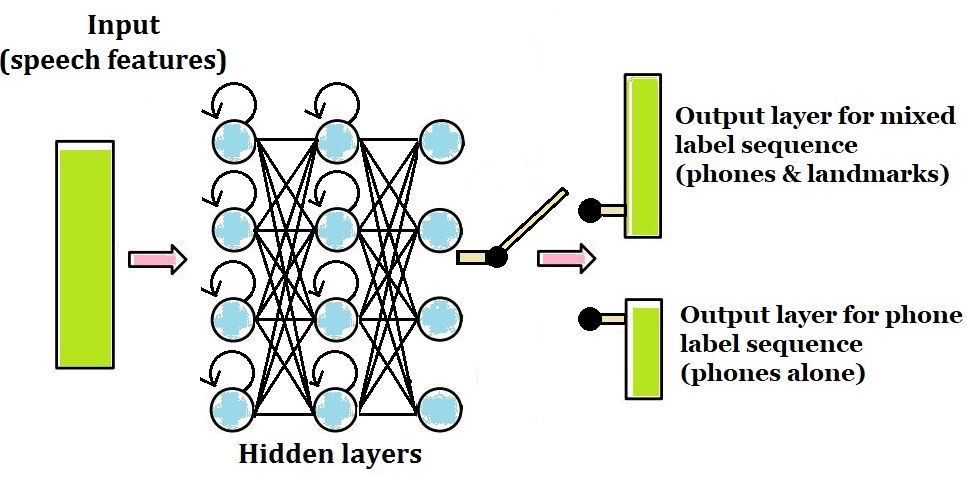}
    \vspace{-4mm}
    \caption{Two-phase acoustic modeling: top output layer pretrains with mixed labels and bottom output layer finetunes with phone labels only}
    \label{fig:mtl}
    \vspace{-0.5cm}
\end{figure}

We follow a pretraining and finetuning procedure to train our CTC models. At the phase of pretraining, the AM initializes weights randomly and is trained by one of our mixed label sequences until convergence; at the phase of finetuning, the AM initializes weights from the pretrained model and continues to be trained by a label sequence with only phones. These two phases of training take the same acoustic features. Figure~\ref{fig:mtl} briefly illustrates the whole procedure. The top output layer calculates a posterior distribution over symbols combined with both phones and landmarks, while the bottom output layer calculates it over only phones. 

\section{Experiments}
\label{sec:result}
\subsection{Configurations}

We conducted our experiments on both the TIMIT~\cite{garofalo1993darpa} and WSJ~\cite{paul1992design} corpora.
We used $40$-dimensional log mel filterbank energy features computed with $10$ms shift and $20$ms span. No delta features or frame stacking were used.
The recurrent neural networks stacked two layers of bidirectional LSTMs, each with 1024 cells (512 cells per direction), capped by a fully connected layer with 256 neurons. Weights are initialized randomly from Xavier uniform distribution~\cite{glorot2010understanding}.
New-Bob annealing~\cite{bourlard2012connectionist} is used for early stopping after a minimum waiting period of two epochs. The initial learning rate is $0.0005$.
The TIMIT baseline is trained on $61$ phones. 
The WSJ baseline is trained on $39$ phones\footnote{\url{https://github.com/Alexir/CMUdict/blob/master/cmudict-0.7b.phones}} defined in the CMU pronunciation dictionary.
One-best greedy search is applied to calculate the phone error rate (PER). We did not map TIMIT phones to CMU phone set (39 phones). In order to make a fair comparison, all baselines went through the same two-phase training with pretraining and finetuning. One-best beam search based on WFSTs is applied to calculate the word error rate in WSJ experiments using  decoding graphs with a primitive trigram (tg) and pruned trigram (tgpr) from EESEN\footnote{\url{https://github.com/srvk/eesen/blob/master/asr_egs/wsj/run_ctc_phn.sh}}. We use the same train/dev/test split from Kaldi Recipes for TIMIT and WSJ.



\subsection{Experiments on TIMIT}
\begin{figure}[htbp]
  \centering
    \includegraphics[width=\linewidth]{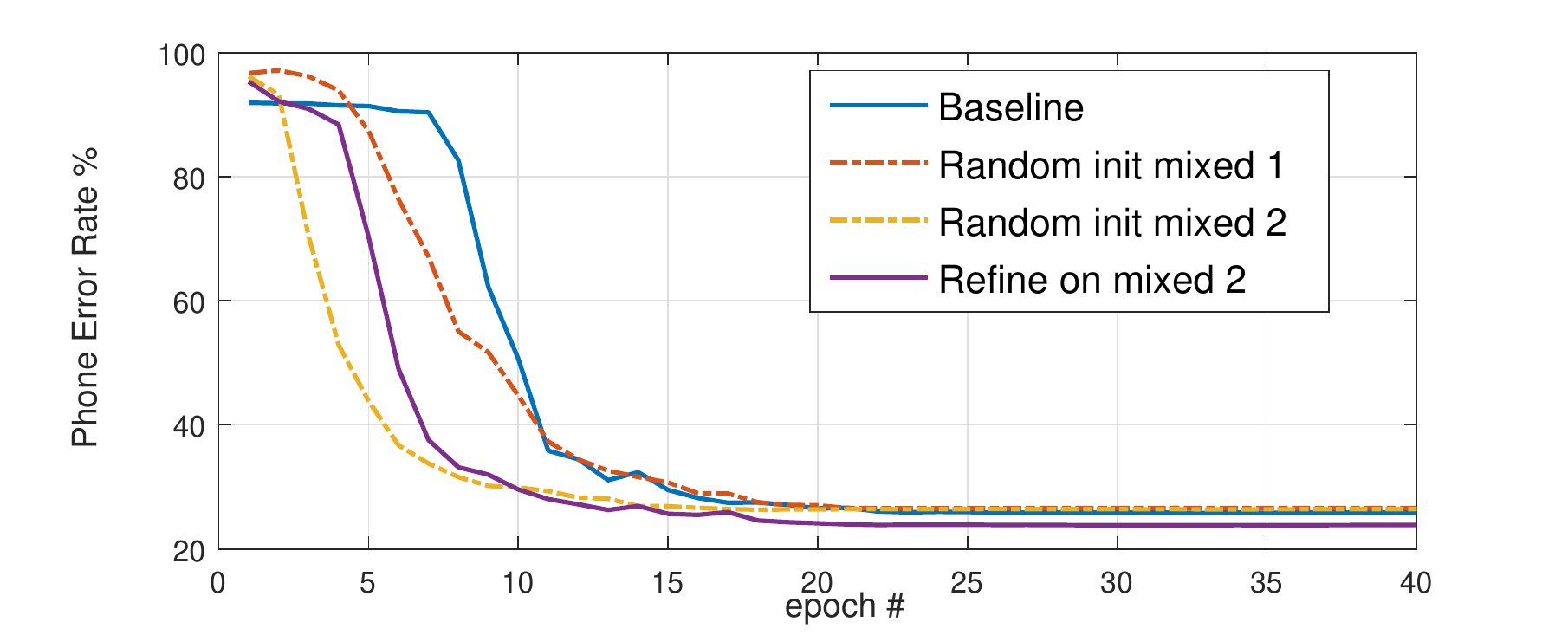}
    \vspace{-8mm}
    \caption{PER as a function of training epoch. PER is calculated against only phones after landmarks are removed.}
    \label{fig:converg}
\end{figure}

Figure~\ref{fig:converg} presents the development set PER as a function of training epoch.
The PER for mixed sequence represented by the red and yellow lines in Figure~\ref{fig:converg} is calculated after landmark labels have been removed from the output sequence. In the pretrain phase, models trained on augmented labels do not seem to have any advantage in terms of error rate. However, the models converge much more rapidly and smoothly. After pretraining, both the baseline and mixed-label systems are finetuned; the mixed-label system (purple line in Fig.~\ref{fig:converg}) 
returns a model that is more accurate.

The exact PERs for different setups on the TIMIT test set are reported in Table~\ref{tab:ler}. Our baseline achieved a PER of $30.36\%$, which was not improved by finetuning.  This is higher than PER reported elsewhere (e.g.,~\cite{graves2006connectionist}), because nobody else calculates PER on the full TIMIT set of 61 phones. 
As shown in Table~\ref{tab:ler}, if we train with mixed labels and strip away landmarks from the hypothesis sequence, landmarks 
provide little benefit.
However, the {\em Mixed 1} and {\em Mixed 2} systems achieved lower PER after the finetuning stage by $4.64\%$ and $8.72\%$ relative, respectively. 
Apparently, a phone sequence augmented with landmarks can be learned more accurately than a raw phone sequence, perhaps because the acoustic features of manner transitions are easy to learn, and help to time-align the training corpus. The {\em Mixed Label 2} set outperforms {\em Mixed Label 1}, apparently because the extra boundary information in {\em Mixed Label 2} is useful to the training algorithm.
\begin{table}[htbp]
\centering
\caption{Comparison between baseline and our proposed models with augmented target labels in PER (\%). Number in the parentheses denotes the relative reduction over baseline.}
\label{tab:ler}
\vspace{2mm}
\begin{tabular}{|l|l|l|l|}
\hline
            & Baseline & Mixed 1     & Mixed 2     \\ \hline
random init & 30.36     & 30.98          & 29.10          \\ \hline
finetuned   & 30.36  & \textbf{28.96} (4.64\%) & \textbf{27.72} (8.72\%) \\ \hline
\end{tabular}
\end{table}

\begin{figure}[htbp]
    \centering
    \includegraphics[width=\linewidth]{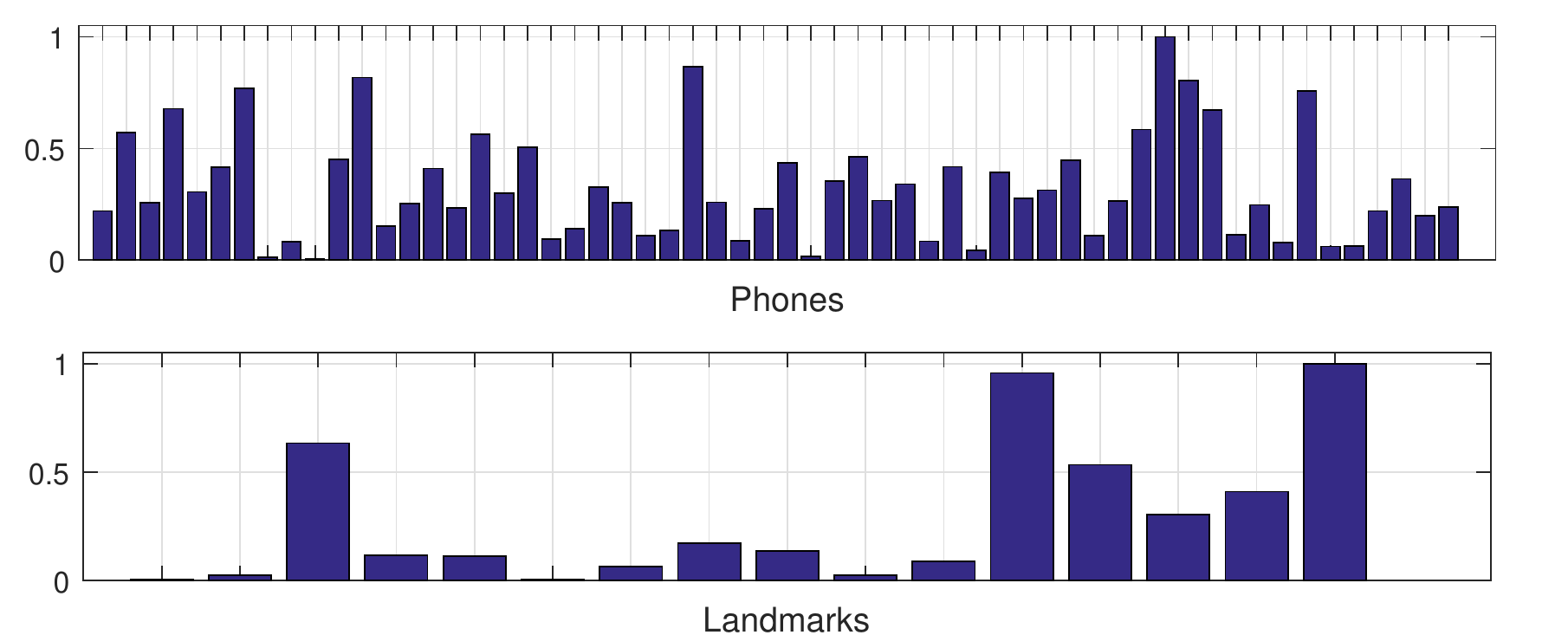}
    \vspace{-7mm}
    \caption{Prior distributions of phones and acoustic landmarks.}
    \label{fig:prior_dist}
\end{figure}

It is not clear why a finetuning stage is needed in order for {\em Mixed 1} to beat the baseline.
One possibility is that landmark labels are helpful for some tokens, and harmful for others; pretraining uses the helpful landmarks to learn better phone alignments, then finetuning permits the network to learn to ignore the harmful landmark tokens.
We looked into the prior distribution on TIMIT, presented in Figure~\ref{fig:prior_dist}, of both phones (top subplot, with phones ordered in the same way as they occurred in Table~\ref{tab:label_mark}) and landmarks (bottom subplot, {\em Mixed Label 2} ordered in category permutation using \emph{continuant} as the first variable and \emph{sonorant} as the second).
The table reveals that the distribution of landmarks is not balanced. Most labels indicate a transition related to the [\emph{+continuant,+sonorant}] phones. A skewed landmark support is not ideal for augmenting phone recognizer training as it tends to provide the same and redundant information for many training sequences.


\subsection{Datasets Smaller and Larger than TIMIT}


To solidify our findings, we further investigated the sensitivity of our approaches to the size of training data on subsets of TIMIT (smaller corpora) and WSJ (a larger corpus). 
In this section, we only demonstrate the experiments using \emph{Mixed Label 2} augmentation method
since it outperforms \emph{Mixed Label 1} in the previous discussion. We report PER/WER results for finetuned models.

\begin{figure}[htbp]
    \centering
    \includegraphics[width=0.9\linewidth]{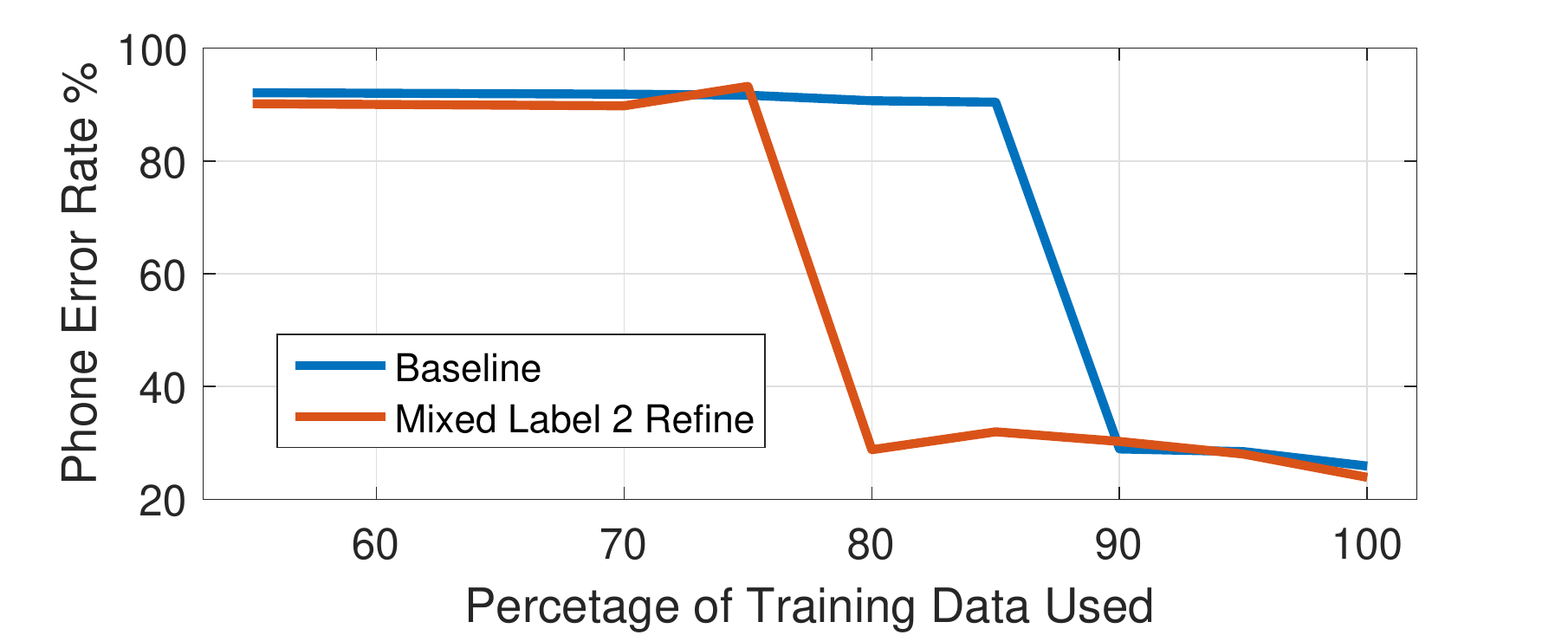}
    \vspace{-4mm}
    \caption{PERs by stretching the amount of training data on TIMIT.}
    \label{fig:per_vary}
\end{figure}

Figure~\ref{fig:per_vary} shows the PER results by stretching the amount of training data on TIMIT.
Both the proposed model and baseline fail to converge when 75\% of the training data is used. We observe that both models start to predict a constant sequence (usually made up of two to three most frequent phones) for all utterances. Scheduled reducing the learning rate by New-Bob annealing can't help to converge to an optimal. Increasing the amount of training data helps both models converge.
The baseline needs 90\% of TIMIT to converge, while the proposed system only needs 80\% of TIMIT.

When scaling up to a even larger corpus on WSJ, the proposed \emph{Mixed Label 2} system could achieve better performance over the baseline consistently in terms of all metrics as shown in Table~\ref{tab:wsj}. Our baseline system slightly under-performs the results published in EESEN~\cite{miao2015eesen} because our network is shallower and the acoustic inputs do not include any dynamic (delta) features, but the benefit of the proposed landmark augmentation method still applies.
To our knowledge, this is the first work to show that manner-change acoustic landmarks reduce both PER and WER on a mid-sized ASR corpus.

\begin{table}[htbp]
\caption{Label Error Rate (\%) on WSJ, where tg and tgpr denote decoding graphs with primitive and pruned trigrams.}
\label{tab:wsj}
\vspace{2mm}
\begin{tabular}{|l|l|l|l|l|}
\hline
   \multirow{2}{*}{} & \multicolumn{2}{c|}{PER} & \multicolumn{2}{c|}{WER ( tgpr / tg )} \\ \cline{2-5}
                     & \multicolumn{1}{c|}{eval92}      & \multicolumn{1}{c|}{dev93}      & \multicolumn{1}{c|}{eval92}          & \multicolumn{1}{c|}{dev93}            \\ \hline
Baseline             & 8.7         & 12.38      & 8.75/8.17       & 13.15/12.31      \\ \hline
Mixed 2            & \textbf{8.12}        & \textbf{11.49}      & \textbf{8.35}/8.19       & \textbf{12.86}/\textbf{12.28}      \\ \hline
\end{tabular}
\end{table}

\section{Conclusion}
\label{sec:conc}

We proposed to augment CTC with acoustic landmarks. We modified the classic landmark definition to suit the CTC criterion and implemented a pretraining-finetuning training procedure to improve CTC AMs. Experiments on TIMIT and WSJ demonstrated that CTC training becomes more stable and rapid when phone label sequences are augmented by landmarks, and achieves a significantly lower (8.72\% relative reduction) asymptotic PER. The advantage is consistent across corpora (TIMIT, WSJ) and across metrics (PER, WER). CTC with landmarks converges when the dataset is too small to train the baseline, and it also converges without the need of time alignments on a mid-sized standard ASR training corpus (WSJ).


\thanks{{\bf Acknowledgements}:
The fifth author was supported by the DARPA LORELEI program.  All results and conclusions are those of the authors, and are not endorsed by DARPA.}
\bibliographystyle{IEEEbib}
\small{\bibliography{refs}}

\end{document}